\documentclass[twocolumn, aps, prl, letterpaper, notitlepage, superscriptaddress]{revtex4-1}

\usepackage{amsmath,amssymb}
\usepackage{graphicx}
\usepackage{import}
\usepackage{color} 
\usepackage{comment}
\definecolor{LinkColor}{rgb}{0,0,.5}

\newcommand{\ket}[1]{|#1\rangle}
\newcommand{\bra}[1]{\langle #1 |}

\newcommand{\tr}[1]{\text{Tr}\{#1\}}

\newcommand{\ubar}[1]{\underline{#1}}

\newcommand{\Ham}{\mathcal{H}}

\begin{document}
\title{Identification and control of electron-nuclear spin defects in diamond}

\author{Alexandre Cooper}
\affiliation{Department of Nuclear Science and Engineering and Research Lab of Electronics,\\Massachusetts Institute of Technology, Cambridge, MA 02139, USA }
\affiliation{Department of Physics, Mathematics and Astronomy,\\California Institute of Technology, Pasadena, CA 91125, USA }

\author{Won Kyu Calvin Sun}
\author{Jean-Christophe Jaskula}
\author{Paola Cappellaro}\email{pcappell@mit.edu}
\affiliation{Department of Nuclear Science and Engineering  and Research Lab of Electronics,\\Massachusetts Institute of Technology, Cambridge, MA 02139, USA }

\date{\today}

\begin{abstract}
We experimentally demonstrate an approach to scale up quantum devices by harnessing spin defects in the environment of a quantum probe. We follow this approach to identify, locate, and control two electron-nuclear spin defects in the environment of a single nitrogen-vacancy center in diamond. By performing spectroscopy at various orientations of the magnetic field, we extract the unknown parameters of the hyperfine and dipolar interaction tensors, which we use to locate the two spin defects and design control sequences to initialize, manipulate, and readout their quantum state. We finally create quantum coherence among the three electron spins, paving the way for the creation of genuine tripartite entanglement. This approach will be useful to assemble multi-spin quantum registers for applications in quantum sensing and quantum information processing.
\end{abstract}
\maketitle

\begin{figure}[ht]
\centering
\includegraphics[width=\columnwidth]{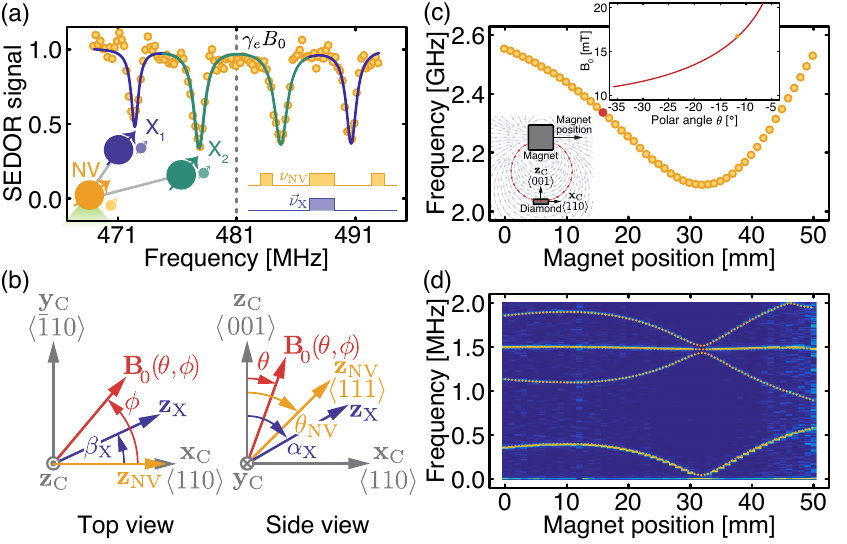}
\caption{
\textbf{Identifying two unknown spin defects in diamond.}
\textbf{(a)}~A single nitrogen-vacancy center (NV) interacts with two electron-nuclear spin defects ($\text{X}_1$, $\text{X}_2$) in diamond. The spin-echo double-resonance (SEDOR) spectrum measured by applying a recoupling $\pi$ pulse at a variable frequency $\vec{\nu}_{\text{X}}$ during a spin-echo on the NV electron spin exhibits two resolved hyperfine doublets centered around the free-electron spin resonance frequency $\nu_e=\gamma_eB_0$, where $\gamma_e$ is the gyromagnetic ratio of the free electron and $B_0=171.8~\text{G}$ is the strength of the static magnetic field oriented along the molecular axis of the NV center. The solid line is a fit to four Lorentzian spectral lines associated with the two hyperfine resonances of $\text{X}_1$ (blue, outer spectral lines) and $\text{X}_2$ (green, inner spectral lines)  
\textbf{(b)}~The unknown parameters of the hyperfine and dipolar tensors of the X spins are measured by varying the strength and orientation of the magnetic field using a permanent magnet. 
The polar and azimuthal angles parametrizing the orientation of the magnetic field ($\theta$, $\phi$) and the principal axes of the hyperfine tensors for the $\text{NV}$ center ($\theta_\text{NV}=54.7^{\circ}$, $\phi_\text{NV}=0^{\circ}$) and $\text{X}$ spins ($\alpha_\text{X}$, $\beta_\text{X}$) are defined with respect to the crystallographic axes ($\bold{x}_C,\bold{y}_C,\bold{z}_C$) of the diamond crystal. 
\textbf{(c)}~NV resonance frequency for various magnet positions. For each magnet position, there exist multiple values of the strength and orientation of the magnetic field that result in the same NV resonance frequency (inset).
\textbf{(d)}~Electron-spin-echo envelope modulation (ESEEM) spectroscopy of the $\text{NV}$ center for various magnet positions. The spectral components at the nuclear frequencies result from hyperfine mixing with the host $^{15}\text{N}$ nuclear spin in the presence of a non-axial magnetic field. For each magnet position, the field strength ($B_0$) and polar angle ($\theta$) are unambiguously determined by finding the simulated spectrum (dotted lines) that best matches the measured spectrum. The frequency wrapping for large magnet positions is an artifact of bandlimited sampling that is captured by our model.
}
\label{Fig1}
\end{figure}

Quantum devices that exploit the spins of impurity atoms or defect sites in solid-state materials offer promising applications in quantum communication~\cite{Childress13, Gao15, Atature18}, quantum information processing~\cite{Morton11, Awschalom13, Watson17} and quantum sensing~\cite{Degen17, Casola18}. Color centers with robust optical transitions and long-lived spin degrees of freedom are especially attractive to engineer optical networks of quantum registers~\cite{Kimble08, Bernien13, Humphreys18} and atomic-scale sensors of time-varying magnetic fields~\cite{Taylor08, Cooper14, Ajoy15}. The most studied of such color centers is the nitrogen-vacancy (NV) center in diamond, because of its outstanding optical and spin properties under ambient conditions~\cite{Doherty13}. 

An important problem with building scalable quantum devices based on synthetic NV centers is the existence of environmental spin defects, mostly byproducts of the NV creation process, such as nitrogen-related centers, vacancies, and their aggregates~\cite{Deak14}. Whereas these spin defects usually cause fluctuations responsible for decoherence~\cite{DeLange10}, they could rather serve as quantum resources were their spin properties under control~\cite{Schlipfe17,Goldstein11}. Although spin defects in the environment of a single NV center have been studied and controlled~\cite{Epstein05, Hanson06, Shi13, Sushkov14, Grinolds14, Knowles16, Rosenfeld18}, a systematic approach to convert electron-nuclear spin defects into useful quantum resources is still needed, e.g., to transfer information between distant quantum registers~\cite{Neumann10, Cappellaro11, Yao11} or improve the sensitivity of quantum sensors~\cite{Goldstein11, Shi15, Lovchinsky16, Simpson17, Cooper19}.

Here, we experimentally demonstrate an approach to identify, locate, and control electron-nuclear spin defects in the environment of a quantum probe using double electron-electron resonance spectroscopy. Our approach relies on exploiting the non-trivial transformation of the spin Hamiltonian under rotation of the external magnetic field to estimate the parameters of the hyperfine and dipolar tensors, as needed to identify and locate unknown spin defects, as well as design control sequences to initialize, manipulate, and readout their quantum state. As a proof-of-principle demonstration, we spectrally characterize two unknown electron-nuclear spin defects in the environment of a single NV center in diamond and create quantum coherence among the three electron spins. These results demonstrate a further step towards assembling large scale quantum registers using electron spins in solid.



Our experimental system consists of a single NV center interacting through magnetic dipole-dipole interaction with two electron-nuclear spin defects ($\text{X}_1$, $\text{X}_2$) randomly created by implanting $^{15}\text{N}$ ions through nanometer-sized apertures in an isotopically-purified diamond crystal~\cite{SupplementalMaterial}. Each $\text{X}$ spin consists of an electronic spin $S=1/2$ strongly-coupled to a nearby nuclear spin $I=1/2$, giving rise to two resolved hyperfine doublets in the spin-echo double-resonance (SEDOR) spectrum~(Fig.~\ref{Fig1}a)~\cite{deLange12}. Each hyperfine transition can be selectively addressed using resonant microwave pulses with negligible crosstalk.
Interestingly, the X spins are stable under optical illumination, enabling repetitive readout of their quantum state~\cite{Cooper19}. 



Our approach to solving the system identification problem consists in estimating the parameters of the spin Hamiltonian describing each of the two electron-nuclear spin defects, 
\begin{eqnarray}
\Ham(\theta,\phi)=\beta_e~\ubar{B}\cdot\ubar{\ubar{g}}\cdot\ubar{S}
+\ubar{S}\cdot\ubar{\ubar{A}}\cdot\ubar{I}
-g_n\beta_n~\ubar{B}\cdot\ubar{I}\label{eq:staticSpinHamiltonian},
\end{eqnarray}
where $\ubar{B}=\ubar{B}(\theta,\phi)$ is the static magnetic field vector of norm $B_0$, $\ubar{\ubar{A}}$ is the hyperfine interaction tensor, $\ubar{\ubar{g}}$ ($g_n$) is the $g$ tensor ($g$ factor) of the electron (nuclear) spin, and $\beta_e$ ($\beta_n$) is the Bohr (nuclear) magneton (we set $\hbar=1$). 

To extract the energy eigenvalues of $\mathcal{H}(\theta,\phi)$, we perform double-resonance spectroscopy for different orientations of the magnetic field~(Fig.~\ref{Fig1}b) by translating and rotating a permanent magnet with respect to the diamond sample. However, simply measuring the resonance frequency of the NV electron spin on only one of its electronic spin transition, e.g., via cw-ESR in the $m_s\in\{0,-1\}$ manifold~(Fig.~\ref{Fig1}c), is not sufficient to uniquely characterize $\ubar{B}(\theta,\phi)$~\cite{Balasubramanian08, Lee15}; there indeed exists multiple admissible pairs of $(B_0, \theta)$ resulting in the same resonance frequency~(inset of Fig.~\ref{Fig1}c).
To resolve this ambiguity, we measure the frequencies of the electron spin-echo envelope modulation (ESEEM)~\cite{Mims72} caused by the strong dipolar coupling to the $^{15}\text{N}$ nuclear spin~(Fig.~\ref{Fig1}d). When the magnetic field is misaligned with respect to the NV molecular axis ($\langle111\rangle$ crystallographic axis), the energy levels of the NV electron and nuclear spins are mixed, such that the spin-echo signal is modulated at the nuclear frequencies and their combinations, $\{\nu_1,\nu_0,\nu_1\pm\nu_0\}$. Performing a numerical fit to the ESEEM spectrum~\cite{Stoll06}, we unambiguously determine the magnetic field strength and polar angle at each magnet position~\cite{SupplementalMaterial}. 


To estimate the parameters of the hyperfine tensors, we follow an approach akin to tomographic imaging reconstruction. Geometrically, the hyperfine tensor can be represented as an ellipsoid, whose dimensions are given by the principal components, $\{A_x, A_y, A_z\}$, and principal angles, $\{\alpha_X,\beta_X, \gamma_X\}$, of the hyperfine tensor. Rotating the magnetic field around a fixed axis generates multiple tomographic cuts of the ellipsoid from which the hyperfine parameters can be estimated.

\begin{figure}[tb]
\centering
\includegraphics[width=\columnwidth]{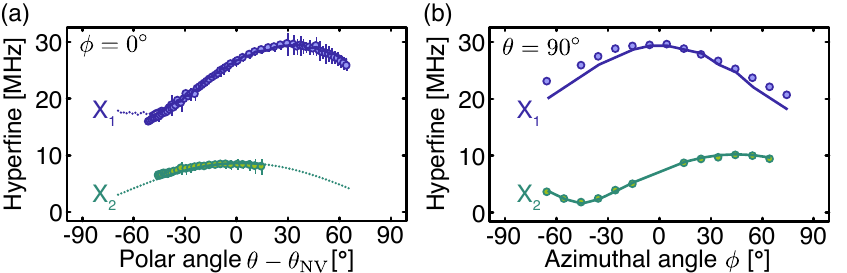}
\caption{
\textbf{Characterizing the hyperfine tensors of two spin defects in diamond.}
{(a)}~Measured hyperfine strengths for various polar angles of the magnetic field ($\theta$) plotted with respect to the polar angle of the NV center ($\theta_{NV}$) in the azimuthal plane $\phi=0^\circ$.
{(b)}~Measured hyperfine strengths for various azimuthal angles of the static magnetic field ($\phi$) in the polar plane $\theta=90^\circ$. The solid lines are the best least-square fit of both sets of data to the eigenvalues of an axially-symmetric hyperfine tensor with four free parameters.}\label{Fig2}
\end{figure}

Specifically, we estimate the hyperfine parameters by monitoring the change in the hyperfine splitting of the $X$ spins as a function of the orientation of the magnetic field~(Fig.~\ref{Fig2}). To simplify the reconstruction problem, we assume the hyperfine tensors $\ubar{\ubar{A}}$ to be axially symmetric ($A_x=A_y\equiv A_\bot$), neglect the nuclear Zeeman term, and choose the $g$ tensor to be isotropic with its principal value equal to the electron spin $g$-factor ($\ubar{\ubar{g}}=g_e\cdot\ubar{\ubar{1}}$). These assumptions are consistent with our measurements, which could be further extended to distinguish between an axially-symmetric tensor and a full tensor~\cite{SupplementalMaterial}. 

Characterizing the hyperfine tensor thus involves measuring a set of four unknown parameters, $\{A_\perp, A_\parallel, \alpha_X, \beta_X\}$, which we experimentally determine by simultaneously fitting the measured hyperfine strengths to the parametric equations for the eigenvalues of $\Ham(\theta,\phi)$. We thus obtain $A_\perp=17.2(3)$, $A_\parallel=29.4(2)$, $\alpha_X=0(2)$, $\beta_X=87(2)$ for $X_1$ and $A_\perp=1.6(3)$, $A_\parallel=11.2(2)$, $\alpha_X=45(2)$, $\beta_X=66(2)$ for $X_2$. We did not find any defects sharing these parameters in the literature~\cite{Loubser78, Isoya92, vanWyk95, Felton09, Liggins10, Atumi13, Green15}, suggesting that they may never have been detected using conventional spectroscopy methods. These defects are possibly nitrogen- or silicon-related centers resulting from nitrogen-ion implantation through a 10-nm amorphous $\text{Si}\text{O}_2$ layer introduced to mitigate ion channeling~\cite{Toyli10}. Further triple-resonance measurements on the X nuclear spins should enable unambiguous identification of the nuclear spin species. 

\begin{figure}[t!]
\centering
\includegraphics[width=\columnwidth]{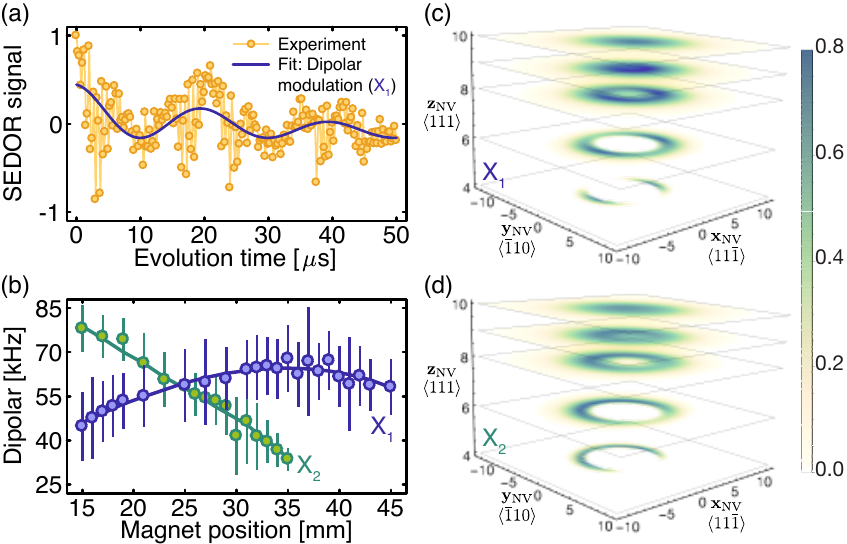}
\caption{
\textbf{Locating two spin defects in diamond.}
{(a)}~Typical dipolar oscillations measured using a recoupled spin-echo sequence in the presence of a non-axial magnetic field. The slow modulation (solid blue line) is caused by the dipolar interaction between the NV electron spin and the $\text{X}_1$ electron spin, whereas the fast modulation is caused by the hyperfine mixing with the NV nuclear spin.
{(b)}~Measured dipolar coupling strengths between the NV electron spin and the $\text{X}_1$ (blue) and $\text{X}_2$ (green) electron spins for various magnet positions. The solid line is the best least-square fit to the eigenvalues of the interacting spin Hamiltonian with three free parameters ($r,\zeta,\xi$, see~\cite{SupplementalMaterial}), which parametrize the relative position of the two X spins with respect to the NV center.
{(c-d)}~Probability distribution maps of the location of the $\text{X}_1$ (top) and $\text{X}_2$ (bottom) spins defined with respect to the coordinate frame of the NV center placed at the origin. The darker color indicates a higher probability of finding the X spin at this specific location. 
}
\label{Fig3}
\end{figure}

To spatially locate the two defects, we measure the change in dipolar interaction strengths as we rotate the polar angle $\theta$ of the magnetic field in the azimuthal plane $\phi=0$. Because the NV and X electron spins are quantized along different axes, the transformation of the dipolar interaction tensor under rotation of the magnetic field is non-trivial~\cite{SupplementalMaterial}. Another complication is that, as the magnetic field is rotated away from the NV molecular axis, the NV coherence signal becomes modulated by the hyperfine interaction with the $^{15}\text{N}$ nuclear spin, in addition to the desired modulation due to the dipolar interaction with the recoupled X electron spin~(Fig.~\ref{Fig3}a).

To generate probability distribution maps for the location of the $\text{X}$ spins with respect to the NV center~(Fig.~\ref{Fig3}c-d), we evaluate the least-square error between the dipolar strengths computed for various admissible locations of the defects and the dipolar strengths measured from the low-frequency components of the SEDOR signal~(Fig.~\ref{Fig3}b). At the most probable location, we estimate the distance from the NV center to be $r_1=9.23(3)~\text{nm}$ and $r_2=6.58(3)~\text{nm}$ for $\text{X}_1$ and $\text{X}_2$ respectively. We searched for signatures of coherent interaction between $\text{X}_1$ and $\text{X}_2$, but could not resolve any, indicating that the two defects are farther apart to each other than to the NV center. 

\begin{figure}[t!]
\centering
\includegraphics[width=\columnwidth]{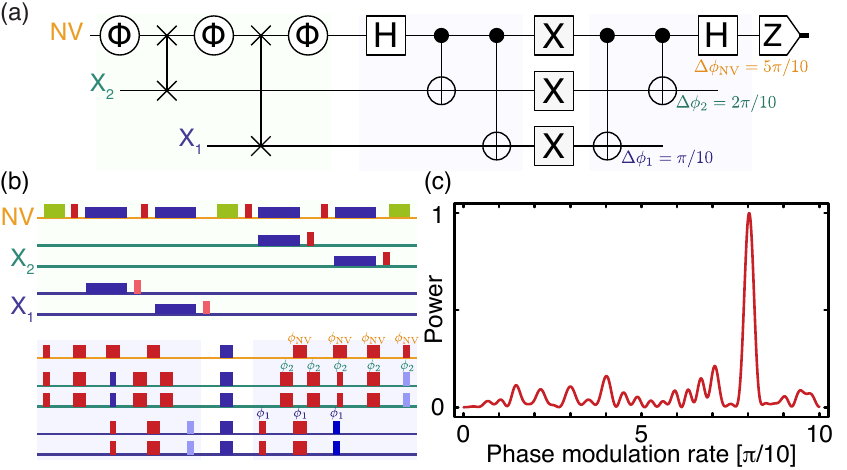}
\caption{\textbf{Creating and detecting three-spin coherence.}
{(a-b)}~The dissipative channel $\Phi$ removes entropy out of the quantum system by optically pumping the NV center (green box). The SWAP operations implemented using Hartmann-Hahn cross-polarization with $\pi/2$ pulses along the $\pm\text{y}$ axis (bright and pale red box) and continuous driving along the $\pm\text{x}$ axis (bright and pale blue box) exchange the state of the NV and X spins, effectively polarizing the X spins. A series of Hadamard (H) and CNOT gates implemented using $\pi/2$ and $\pi$ pulses realize an entangling and disentangling gate to create and detect three-spin coherence, which is protected against dephasing by a series of X gates. The three-spin coherence is mapped back into a population state of the NV spin using a series of disentangling gates and measured projectively in the Z basis (green box). The phase of the pulses of the disentangling gate on the NV, $\text{X}_1$, and $\text{X}_2$ spins are incremented by steps of $\Delta\phi_{\text{NV}}=5\pi/10$, $\Delta\phi_2=2\pi/10$, and $\Delta\phi_1=\pi/10$ respectively to spectrally label the spin coherence terms. {(c)}~The power spectrum of the the signal shows parity oscillations~\cite{Monz11} at the sum of the three modulation rates ($\Delta\phi_\Sigma=8\pi/10$), thus indicating the creation of three-spin coherence~\cite{SupplementalMaterial}.
}\label{Fig4}
\end{figure}

Having partially identified and located the two X defects, we now have a consistent description of the three-spin system that is sufficient to design control protocols to engineer its quantum state and create correlated states of multiple spins. To demonstrate control of the two X spins~\cite{SupplementalMaterial}, we create quantum coherence among the three electron spins, taking a first step towards the creation of genuine tripartite entanglement. Our control protocol (Fig.~\ref{Fig4}) is based on (1) initializing the three-spin system in a pure state using coherent spin-exchange with the optically-polarized NV spin, (2) creating three-spin coherence using a series of entangling CNOT gates, and (3) mapping the coherence back into a population difference on the NV spin using a series of disentangling CNOT gates with modulated phases. 


Specifically, because the X spins lacks a known mechanism for dissipative state preparation, we first initialize its quantum state using multiple rounds of Hartmann-Hahn cross-polarization~\cite{Hartmann62,Belthangady13,Laraoui13}, which relies on simultaneously driving both spins at the same Rabi frequency to engineer coherent spin exchange in the rotating frame at a rate given by the dipolar coupling strength. 

We then create three-spin coherence by synthesizing entangling gates using the recoupled spin-echo sequence~\cite{deLange12}, which decouples the NV spin from its environment using a spin-echo sequence while selectively recoupling the dipolar interaction with the X spin with a recoupling $\pi$ pulse. The recoupled spin-echo sequence correlates the two spins, but does not necessarily create entanglement in the presence of control imperfections.


We finally quantify the amount of three-spin coherence created after the entangling gate by mapping it back into a measurable population difference on the NV spin. To distinguish between the creation of single-spin and multi-spin coherence, we increment the phase of the pulses of the disentangling gates after each realization of the experiment by steps of $5\pi/10$, $2\pi/10$, and $\pi/10$ for the NV, $\text{X}_1$, and $\text{X}_2$ spins respectively. Although this method~\cite{Scherer08,Mehring03} does not provide full state tomography, the modulation of the phases results in a modulation of the polarization signal (Fig.~\ref{Fig4}c) at a rate given by the sum of the three increment rates ($8\pi/10$), thus indicating the creation of three-spin coherence without significant leakage to other coherence terms and showing a first step towards the creation of genuine tripartite entanglement, which could be used to achieve quantum-enhanced sensing~\cite{Cooper19}.


In conclusion, we have demonstrated an approach to identify and control electron-nuclear spin defects in the environment of a quantum probe using double-resonance spectroscopy. This approach will be useful to characterize unknown spin defects in solids so as to better understand their formation mechanisms, mitigate their detrimental influence, and harness their favorable spin, charge, and optical properties. This approach will also be useful to identify spin systems of greater complexity, including unknown molecular structures placed near the surface of diamond~\cite{Shi15, Lovchinsky16}. Instead of more abundant species such as substitutional nitrogen defects (P1 centers) and free electrons, harnessing proximal electron-nuclear spin defects will enable better spectral separability in crowded spectrum and better stability against photoionization. Controlling their nuclear spins, e.g., by direct driving at the Larmor frequency using radio-frequency pulses in triple-resonance experiments, will provide further access to quantum resources to process and store quantum information.

\begin{acknowledgments}
This work was in part supported by NSF grants PHY1415345 and EECS1702716. A.~C.  acknowledges financial support by the Fulbright Program and the Natural Sciences and Engineering Research Council of Canada. We are grateful to Chinmay Belthangady and Huiliang Zhang for their experimental support.
\end{acknowledgments}


%

\pagebreak
\onecolumngrid
\begin{center}
\textbf{Supplemental Material\\Identification and Control of Electron-Nuclear Spin Defects in Diamond}
\end{center}

\setcounter{equation}{0}
\setlength{\parskip}{0.2cm}
\section{Preparing the diamond sample}
The nitrogen-vacancy (NV) center in diamond is a point-defect formed by a substitutional nitrogen atom located nearby a vacancy in the diamond lattice. The NV center in its negatively charged state ($\text{NV}^{-}$) has two unpaired electrons that form a spin-triplet ground state with three magnetic sub-levels, $m_s=\{0, \pm1\}$, and a zero-field splitting of $\Delta=2\pi\cdot2870~\text{MHz}$. The NV center has a spin-triplet excited state with a phonon-broaden spin-preserving optical transition in the visible range centered at $637~\text{nm}$. This optical transition enables spin-state initialization in the $m_s=0$ ground state by optical pumping and spin-state readout by fluorescence imaging. The  host nuclear spin ($I=1$ for N-14, $I=1/2$ for $^{15}\text{N}$), coupled by an anisotropic hyperfine interaction, provides additional degrees of freedom for storing quantum information and assisting in magnetic sensing applications.

We fabricated two-dimensional arrays of confined ensembles of spin defects in a synthetic diamond crystal~\cite{Toyli10}, by implanting $^{15}\text{N}$ nitrogen ions through circular apertures with a diameter of $30~\text{nm}$. The diamond substrate was a single crystal chemical vapor deposition (CVD) diamond from Element Six with a $100~\mu\text{m}$-thick layer of isotopically enriched $99.999~\%$ $^{12}$C grown on top of a $300~\mu\text{m}$-thick electron grade single crystal diamond substrate. The diamond sample was cut with its edge directed along the $\langle110\rangle$ crystallographic axis, such that the $\langle111\rangle$ molecular axis of the NV center lied in the $\langle110\rangle\times\langle001\rangle$ crystallographic plane with its transverse projection oriented towards the $\langle110\rangle$ edge of the diamond sample.

After cleaning the surface with boiling acid, we deposited a 10-nm $\text{Si}\text{O}_{2}$ layer to mitigate ion channeling during ion implantation.
 We  then coated the sample with a $150~\text{nm}$-thick layer of Poly(methyl methacrylate) (PMMA) resist and thermally evaporated Au. We used  electron-beam lithography with an exposure dose of $1400~\mu\text{C}/\text{cm}^2$ to pattern nano-aperture arrays, and finally developed the PMMA resist while keeping the SiO2 layer. 

We then implanted $^{15}\text{N}$ nitrogen ions with an energy of $14~\text{keV}$ and a dose of $10
^{13}~\text{cm}^{-2}$. We chose these implantation energy and dose parameters as a trade-off between increasing the mean distance to the diamond surface and reducing the longitudinal straggling of the nitrogen ions. We further annealed the diamond sample at a temperature of $800^\circ\text{C}$ for $4~\text{h}$ to promote the mobility of vacancies and create NV centers with a conversion efficiency of less than a few percent. We finally cleaned the surface of the diamond with a boiling mixture of concentrated acids (1:1:1 $\text{H}_2\text{SO}_4:\text{HNO}_3:\text{HClO}_4$). We routinely cleaned the diamond surface with a piranha acid solution (3:1 $\text{H}_2\text{O}_2:\text{H}_2\text{SO}_4$) and did not observe any modifications of the properties of our spin system.

Numerical simulations with the SRIM software indicated that the spatial distribution of substitutional nitrogen defects in each implanted region was normally distributed with a mean implantation depth of $19.9~\text{nm}$ and a longitudinal straggling of $6.6~\text{nm}$, greater than the interaction range with surface spins. 
We searched over more than 150 implanted regions to identify three single NV centers, one of which exhibited a strongly modulated interferometric signal and was thus used in this study. For this NV center, we could not resolve a coherent signal from ensembles of nuclear spins associated with impurities on the surface of the diamond sample or protons of the confocal oil, suggesting a relatively deep NV center. 

\section{Determining the strength and orientation of the magnetic field}
As the parameters of the dipolar and hyperfine tensors depend critically on the orientation of the static magnetic field, we implemented a precise protocol to extract its strength and orientation from spectroscopic measurements on the NV center. 

To systematically vary the orientation of the static magnetic field at the location of the NV center, we mounted a $25.4~\text{mm}$-edge cubic magnet on a linearly-actuated translation stage with rotational degrees of freedom. We aligned the magnetization axis of the magnet along the $\langle110\rangle$ crystallographic axis of the diamond crystal in such a way that displacing the magnet along its magnetization axis rotated the magnetic field by the polar angle $\theta$ in the $\langle001\rangle\times\langle110\rangle$ ($\phi=0^\circ$) crystallographic plane.  This  was confirmed by performing spectral measurements on an ensemble of NV centers implanted in a nearby region of the same diamond and observing the spectral overlap of the resonance frequencies of two out of the four NV crystallographic classes. Fixing the position of the magnet at the polar angle $\theta=90^\circ$ and rotating the magnet along its vertical axis further rotated the azimuthal angle $\phi$ of the magnetic field in the $\langle110\rangle\times\langle\bar{1}10\rangle$ crystallographic plane.

\begin{figure*}[htb]
\centering
\includegraphics[width=0.9\textwidth]{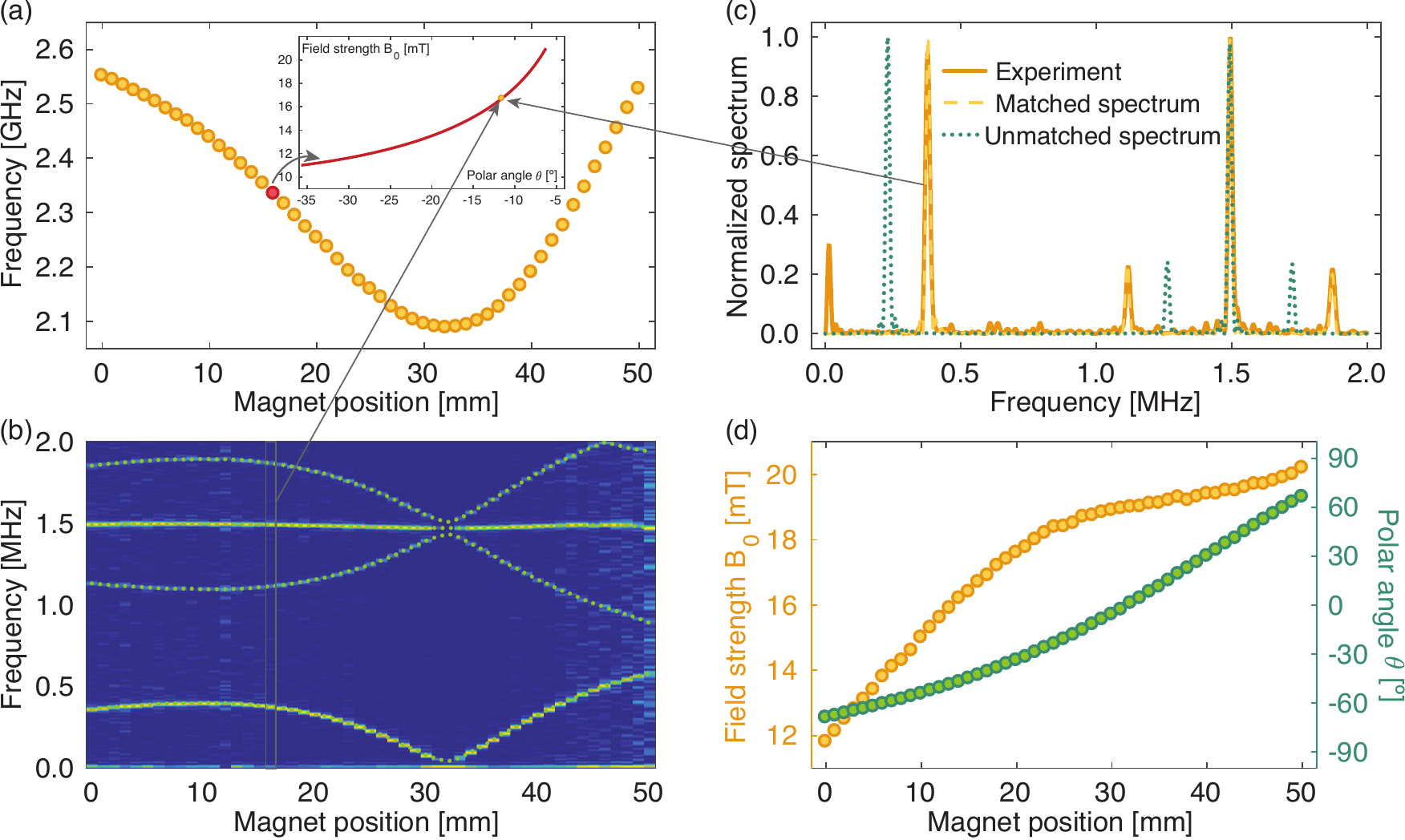}
\caption{
\textbf{Measuring the strength and orientation of the static magnetic field.}
{(a)}~Measurements of the resonance frequency of the NV electron spin for various magnet positions. For each magnet position, there exist multiple values of the strength and orientation of the magnetic field that result in the same NV resonance frequency (inset).
{(b)}~Measurements of the electron spin-echo envelope modulation (ESEEM) of the NV electron spin for various magnet positions. The spectral lines at the nuclear frequencies result from hyperfine mixing with the host \mbox{$^{15}\text{N}$} nuclear spin in the presence of a non-axial magnetic field.
{(c)}~For each magnet position, the field strength and polar angle are unambiguously determined by finding the simulated spectrum that best matches the measured spectrum.
{(d)}~Field strength and polar angle of the magnetic field recovered from a series of spectral measurements on the NV electron spin.
}
\label{Fig2SM}
\end{figure*}

To quantify the strength and orientation of the static magnetic field at each position of the magnet, we first measured the resonance frequency of the NV electron spin in the $m_s\in\{0,-1\}$ manifold using a continuous-wave electron spin resonance (cw-ESR) sequence~(Fig.~\ref{Fig2SM}a). The set of measured frequencies was, however, not sufficient to uniquely determine the strength and orientation of the static magnetic field; there indeed existed an infinite number of admissible values for $(B_0, \theta)$ that resulted in the same resonance frequency~(inset of Fig.~\ref{Fig2SM}a).  

To resolve this ambiguity, we further measured the frequencies of the electron spin-echo envelope modulation (ESEEM)~\cite{Mims72} caused by the strong dipolar coupling to the intrinsic $^{15}\text{N}$ nuclear spin of the NV center. 
As the hyperfine tensor of the $^{15}\text{N}$ spin is axially symmetric, no ESEEM is detectable when the static magnetic field is oriented along the molecular axis of the NV center ($\langle111\rangle$ crystallographic axis). 
For any other orientation of the magnetic field ($\theta'=\theta-\theta_{NV}\neq0$), the bare energy levels of the NV electron spin and \mbox{$^{15}\text{N}$} nuclear spin are mixed, resulting in an effective interaction strength proportional to $A_\perp B\sin(\theta')/(\Delta-\gamma_eB\cos(\theta'))$.
Then, the spin-echo signal is modulated at the nuclear frequencies and their combinations, $\{\nu_1,\nu_0,\nu_1\pm\nu_0\}$. These frequencies correspond to the quantization energies of the nuclear spin conditional on the NV electron spin being in the $m_s=0$ or $m_s=-1$ spin states, including the mixing contributions. 

We numerically simulated the ESEEM spectrum~\cite{Stoll06} by diagonalizing the electron-nuclear spin Hamiltonian of the NV center for different values of the strength and orientation of the static magnetic field. For each position of the magnet, we searched for the field parameters $(B_0,\theta)$ that best reproduced the measured ESEEM spectrum under the constraint of a known NV resonance frequency. Following this approach, we determined a unique pair of admissible values for the field parameters at each magnet position~(inset of Fig.~\ref{Fig2SM}d).  

\section{Characterizing the hyperfine interaction strength}
We derive an analytical expression for the hyperfine coupling strength in the secular approximation as a function of the orientation of the static magnetic field. The Zeeman Hamiltonian for the electron spin is given by
\begin{eqnarray}
\Ham_e(\theta,\phi)&=&\beta_e~\ubar{B}\cdot\ubar{\ubar{g}}\cdot\ubar{S}\\
~&=&g_e \beta_e ~\ubar{B}\cdot\ubar{S}\\
~&=&\omega_e (\cos(\theta)S_z+\sin(\theta)(\cos(\phi)S_x+\sin(\phi)S_y)),
\end{eqnarray}
where $\omega_e = g_e \beta_e B_0$ is the Zeeman energy of the electron spin and
\begin{eqnarray}
\ubar{B}(\theta,\phi)=B_0 (\sin(\theta)\cos(\phi), \sin(\theta)\sin(\phi), \cos(\theta)), 
\end{eqnarray}
is the magnetic field vector expressed in the crystal frame using the polar and azimuthal angles $(\theta, \phi)$.

The strength of the magnetic field is chosen such that the hyperfine coupling strength is smaller than the electron spin Zeeman energy, but larger than the nuclear spin Zeeman energy, i.e., $\omega_n\ll \|A\|\ll\omega_e$. Under this assumption, the electron spin is quantized by the Zeeman energy, whereas the nuclear spin is not. 

Let's recall that the hyperfine tensor is fully characterized by its principal components, $\{A_x, A_y, A_z\}$, and its orientation with respect to the crystal frame given by the Euler angles, $\{\alpha,\beta,\gamma\}$. The hyperfine tensor in its principal coordinate frame is thus represented by a diagonal matrix 
\begin{eqnarray}
A=\textrm{diag}[A_x,A_y,A_z],
\end{eqnarray}
which can be rotated into the crystal frame as
\begin{eqnarray}
\Hat{\Hat{A}}=R^T\cdot A\cdot R,
\end{eqnarray} 
where $R$ is the rotation matrix describing the transformation of the hyperfine matrix from its principal coordinate frame to the crystal frame, 
\newcommand{\ca}{\cos(\alpha)}
\newcommand{\sa}{\sin(\alpha)}
\newcommand{\cb}{\cos(\beta)}
\newcommand{\sbe}{\sin(\beta)}
\newcommand{\cg}{\cos(\gamma)}
\newcommand{\sg}{\sin(\gamma)}
\begin{eqnarray*}
  R&=&\left(\begin{array}{rrr}
    \cg\cb\ca-\sg\sa&\cg\cb\sa+\sg\ca&-\cg\sbe\\
    -\sg\cb\ca-\cg\sa&-\sg\cb\sa+\cg\ca&\sg\sbe\\
    \sbe\ca &\sbe\sa &\cb
  \end{array}\right).
\end{eqnarray*}

For simplicity, assume the magnetic field to be aligned along the $z$-axis of the crystal frame, such that $\ubar{B}=B_0\cdot(0,0,1)$ for $\theta=0$ and $\phi=0$. The secular hyperfine Hamiltonian is given by
\begin{eqnarray}
\mathcal{H}_h&=&\vec S\cdot \Hat{\Hat{A}}\cdot \vec I\\
~&\approx&S_z{\Hat{A}_z}\cdot \vec I=S_z(A_{zx}I_x+A_{zy}I_y+A_{zz}I_z),
\end{eqnarray}
giving rise to an effective hyperfine frequency shift of $C_z=\sqrt{A_{zx}^2+A_{zy}^2+A_{zz}^2}$.

In general, the hyperfine coupling strength is given by
\begin{eqnarray}
C_z= \sqrt{\tr{(H_e\otimes I_x) H_h }^2+\tr{(H_e\otimes I_y) H_h }^2+\tr{(H_e\otimes I_z) H_h }^2}/4\omega_e.
\end{eqnarray}
In the case of an isotropic hyperfine tensor, we have
\begin{eqnarray}
C_z^{\text{iso}}=A_z,
\end{eqnarray}
whereas in the case of an axially symmetric tensor, we have
\begin{equation}
\begin{array}{ll}
C_z^{\text{ax}}&=
\frac1{2\sqrt{2}}
\left[5 A_x^2+3 A_z^2-\left(A_x^2-A_z^2\right)\times\right.
\\ &
\left.\left(4 \cos (2 \delta ) \sin ^2(\beta ) \sin ^2(\theta )+
4 \cos ( \delta) \sin (2 \beta ) \sin (2 \theta )+
\cos (2 \beta ) (3 \cos (2 \theta )+1)+\cos (2 \theta )\right)\right]^{1/2},
\end{array}
\label{eq:axialhyperfine}
\end{equation}
where $\delta=\alpha-\phi$.
In the general case of an arbitrary tensor, we can also obtain an explicit expression, which is given by
\small
\renewcommand{\arraystretch}{1.5}
\begin{equation}\begin{array}{ll}
C_z&=\displaystyle \frac{1}{4} \left[
5 \left(A_x^2+A_y^2\right)+6 A_z^2+8 \left(A_x^2-A_y^2\right) \sin (2 \gamma ) \left(\cos (\beta ) \sin (2 \delta  ) \sin ^2(\theta )-\sin (\beta ) \sin (\delta  ) \sin (2 \theta )\right)
\right.\\
&\displaystyle+\left(A_x^2-A_y^2\right) \cos (2 \gamma ) 
\left(2 (\cos (2 \beta )+3) \cos (2 \delta  ) \sin ^2(\theta )-4 \sin (2 \beta ) \cos (\delta  ) \sin (2 \theta )+2 \sin ^2(\beta ) (3 \cos (2 \theta )+1)\right)\\
&\displaystyle\left.
+\left(2 A_z^2-A_x^2-A_y^2\right) \left(4 \sin ^2(\beta ) \cos (2 \delta  ) \sin ^2(\theta )+4 \sin (2 \beta ) \cos (\delta  ) \sin (2 \theta )+\cos (2 \beta ) (3 \cos (2 \theta )+1)+\cos (2 \theta )\right)
\right]^{1/2}.
\end{array}	
\label{eq:fullhyperfine}
\end{equation}
\normalsize

We fit the data collected as described in the main text to these formulas,  where the magnetic field angles $\theta$ and $\phi$ were varied by translating the permanent magnet with respect to the diamond crystal. We found that the model of an axially symmetric tensor (Eq.~(\ref{eq:axialhyperfine})) fitted slightly better the data (lower $\chi^2$) than the general model of Eq.~(\ref{eq:fullhyperfine}). Further data points at different combinations of $\theta$ and $\phi$ could better discriminate between models.

\section{Characterizing the dipolar interaction strength}
In the absence of an external static magnetic field, the NV center is quantized along its $\langle111\rangle$ molecular axis defined by the strong crystal field responsible for the zero-field splitting $\Delta=2\pi\cdot2870~\text{MHz}$. In the presence of a weak static magnetic field of strength $\gamma_eB_0\ll \Delta$, the NV electronic spin is weakly tilted away from its molecular axis, whereas the X electronic spin is predominantly quantized along the external field. This behavior is responsible for a non-trivial transformation of the dipolar interaction tensor under rotation. The dependence of the effective dipolar interaction strength on the orientation of the static magnetic field thus provides information about the spatial location of the X spins with respect to the NV center.

The total spin Hamiltonian describing the interaction between the NV center ($S_\text{NV}=1$, $I_\text{NV}=1/2$) and the X electron-nuclear spin defect ($S_\text{X}=1/2$, $I_\text{X}=1/2$) is given by
\begin{eqnarray}
\Ham = \Ham_{\text{NV}} + \Ham_{\text{X}} + \Ham_{\text{NV}-\text{X}},
\end{eqnarray}
where $\Ham_{\text{NV}}$ ($\Ham_{\text{X}}$) is the spin Hamiltonian of the NV center (X spin defect) and $\Ham_{\text{NV}-\text{X}}$ the interaction Hamiltonian describing the magnetic dipolar interaction between the NV electron spin and X electron spin. The dipolar Hamiltonian in its general form is given by
\begin{eqnarray}
\Ham_{\text{NV}-\text{X}} = -\frac{\mu_0}{4\pi} \frac{\gamma_{\text{NV}}\gamma_{\text{X}}\hbar^2}{r^3} (3(\ubar{S}_{\text{NV}}\cdot\ubar{r})(\ubar{S}_{\text{X}}\cdot\ubar{r})-(\ubar{S}_{\text{NV}}\cdot\ubar{S}_{\text{X}})),
\end{eqnarray}
where $\ubar{r}=(\sin{(\zeta)}\cos{(\xi)}, \sin{(\zeta)}\sin{(\xi)}, \cos{(\zeta)})$ is the interatomic vector of norm 1 that join the NV center and X spin defect, parameterized by the distance $r$ between the two centers and the polar and azimuthal angles $(\zeta,\xi)$ defined with respect to the NV molecular axis.


Since we can consider the dipolar coupling as a perturbation of each spin Hamiltonian, the only visible component in an experimental measurement is the energy-conserving one. 
Thus, the effective (secular) dipolar coupling strength between the NV electron spin and X electron spin is obtained by computing the eigenvalues of the total dipolar  Hamiltonian in the doubly-tilted frame
\begin{eqnarray}
\tilde{\Ham} = U_{\text{X}}^{-1}U_{\text{NV}}^{-1}\Ham_{\text{NV}-\text{X}}U_{\text{NV}}U_{\text{X}}
\end{eqnarray}
where $U_{\text{NV}}$ ($U_{\text{X}}$) is the unitary transformation diagonalizing $\Ham_{\text{NV}}$ ($\Ham_{\text{X}}$).

By projecting the NV electron spin onto an effective two-level system, it is possible to analytically evaluate the secular dipolar strength. In this approximation, valid when the $m_s=+1$ level is energetically isolated and never populated by the driving field, we obtain:
\[d=d_c\frac{3 \sin (2 \zeta ) \cos (\xi) \sin (\theta' ) [\Delta -3 \gamma_eB_0 \cos (\theta' )]-6 \gamma_eB_0 \sin ^2(\zeta ) \cos (2 \xi ) \sin ^2(\theta')+(3 \cos (2 \zeta )+1) (\Delta  \cos (\theta' )-\gamma_eB_0 \cos (2 \theta' ))}{4 r^3 \sqrt{2 (\gamma_eB_0 \sin(\theta' ))^2+(\Delta -\gamma_eB_0 \cos (\theta' ))^2}}\]
where $d_c = 2\pi\cdot52.041~\text{kHz}$ is the dipolar constant for two electronic spins at a distance of $1~\text{nm}$ and $\theta'=\theta-\theta_\text{NV}$ is the angle between the static magnetic field and the NV molecular axis in the $\bold{y}_\text{NV}=0$ plane.

\section{Controlling the three electron-nuclear spin system\\Generation and detection of three-spin coherence}
Identifying the unknown parameters of the three-spin Hamiltonian enables precise control of the three-spin system via resonant microwave pulses and free evolution under dipolar interaction. To demonstrate control, we generated and detected three-spin coherence. The spin system was first initialized by exploiting cross-polarization under continuous driving and optical polarization of the NV system. We performed the cross polarization on each hyperfine transition of the X spin in series, as the hardware implementation was easier to avoid cross-talks. Similarly, we applied microwave pulses on each hyperfine transition in series rather than simultaneously. 

We then used a recoupled spin-echo sequence to generate CNOT gates, combining single-spin $\pi/2$ pulses and free evolution under the spin-spin couplings. Intuitively, the free evolution blocks engineer a controlled-Z rotation, which is transformed into a CNOT by the $\pi/2$ pulses. In order to protect the spins affected by the gates from noise and other couplings in the system, we embedded spin echoes in the gate.

The polarization sequence ideally prepares the state $\ket{000}$. The entangling control then prepares, in the absence of control errors, the GHZ state $\ket{\textrm{GHZ}}=\frac1{\sqrt{2}}(\ket{000}+e^{i\chi}\ket{111})$. We repeat the experiment after incrementing the phase of the detection pulses by steps of $\Delta\phi_{\text{NV}}=5\pi/10$, $\Delta\phi_{X2}=2\pi/10$, and $\Delta\phi_{X1}=\pi/10$ for a total of 64 repetitions. Given these phase increments, only if there exists three-spin coherence, we would measure a modulation of the signal at the sum of the three modulation rate, $\Delta\phi_\Sigma=\Delta\phi_{\text{NV}}+\Delta\phi_{X2}+\Delta\phi_{X1}=8\pi/10$.

We further notice that, given an arbitrary mixed state, assuming the disentangling gate to be perfect, we would expect modulation only at the rates of $0,\pi/10, 2\pi/10, \dots, 8\pi/10$. Thus, our choice of modulation frequencies simplifies the expected modulation spectrum. As shown in the spectrum presented in the main text, most of the signal is indeed given by the  $\Delta\phi_\Sigma=8\pi/10$ component. However, we observe some small leakage at other spectral components ($3\pi/10$, $4\pi/10$, and $7\pi/10$), indicating that the disentangling gate is not perfect. We note that these imperfections would affect any attempts to perform full state tomography, thus preventing us from distinguishing between imperfect state preparation and imperfect state readout. 

The frequency modulation can still give estimates of the prepared state fidelity and entanglement. 
First, we note that our experiment effectively post-selects on the charge state of the NV center: indeed, we consider the difference signal obtained when measuring the NV center in the $\ket{0}$ and $\ket{-1}$ state, so that the signal measured when the NV is ionized in the NV$^0$ state is mainly canceled out as it would typically contribute just a common background noise. 

\begin{figure*}[h]
\centering
\includegraphics[width=0.5\textwidth]{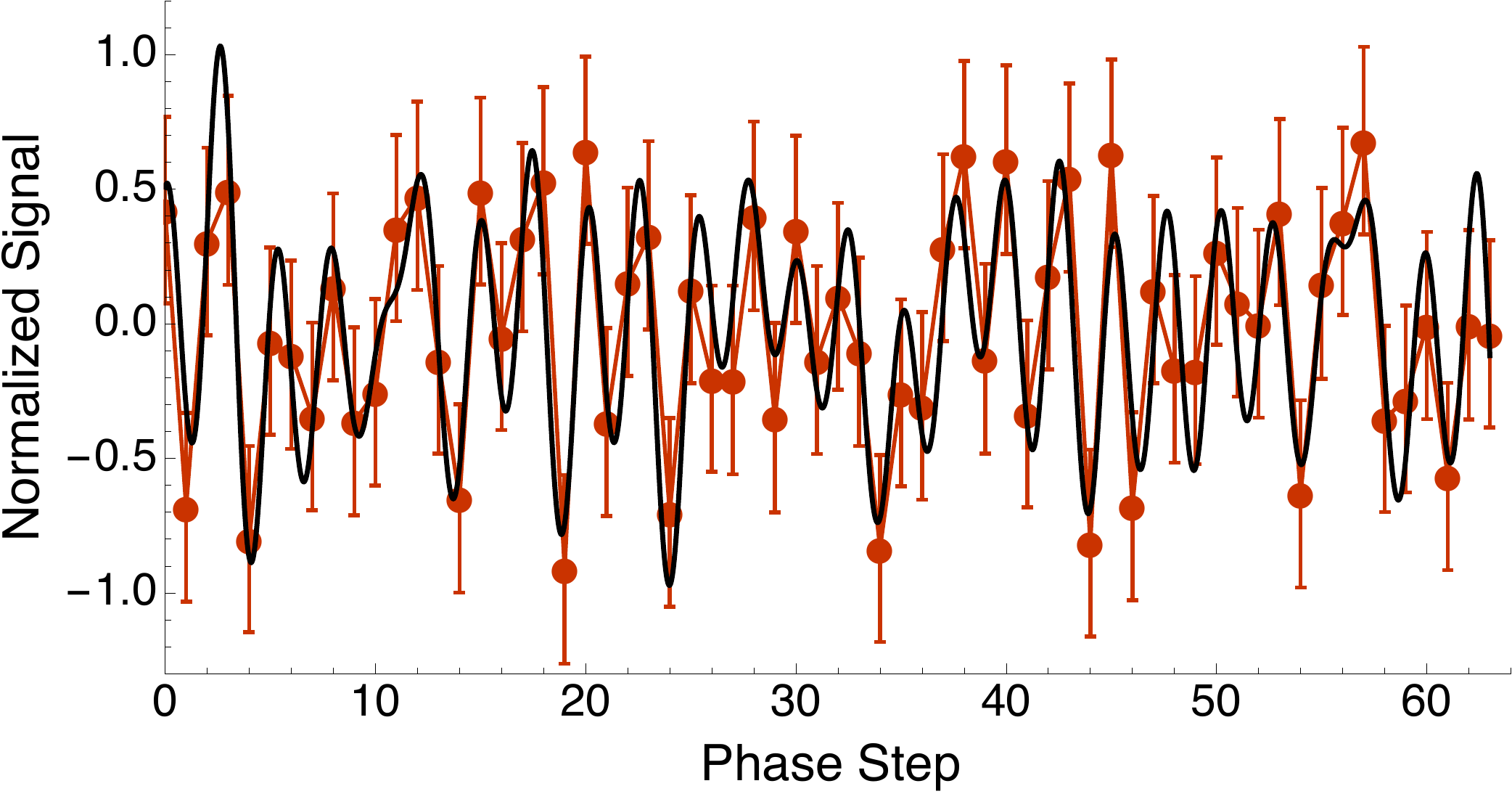}
\caption{
\textbf{Normalized modulation signal.} The solid line is a non-linear fit of the experimental data (circles) to a sum of sinusoids.
}
\label{Fig3SM}
\end{figure*}

We can fit the difference signal to a sum of sinusoids~(Fig.~\ref{Fig3SM}), $S_d=a_0+\sum_i a_i\cos(\phi_i+\omega_i n)$, where $n$ is the step size in the experimental phase increment. The component $a_8$ at $\omega_8=\Delta\phi_\Sigma=8\pi/10$ yields the absolute value of the coherence  $a_8/2=|\rho_{18}|=|\bra{111}\rho\ket{000}|$, where $\rho$ is the state prepared by the series of cross-polarization and entangling gates. By renormalizing the signal by its L$^2$ norm, $\int S_d^2=1/2$, we effectively post-select the initialized state in the subspace spanned by the observable. We then observe $a_8=2|\rho_{18}|=0.43(5)$, indicating that most of the signal is indeed at the desired modulation frequency, showing good coherent control of the three-spin system.


We note that the fidelity is given by $F=\bra{\textrm{GHZ}}\rho\ket{\textrm{GHZ}}=\frac12(\rho_{11}+\rho_{88}+2|\rho_{18}|)$, where we allowed for a free parameter, the phase $\chi$. Also, the state is entangled if we have $\textrm{Tr}[\rho W]<0$, with $W$ an entanglement witness~\cite{Acin01,Rahimi07} given by $W=\frac34-\ket{\textrm{GHZ}}\bra{\textrm{GHZ}}$, that is, there is genuine tripartite entanglement if the fidelity is $F>\frac34$. If the only off-diagonal term in the density matrix were $\rho_{18}$, we can bound $\rho_{11}+\rho_{88}$ by imposing that the density matrix should be semidefinite positive, and the fidelity is bounded by $F\geq2|\rho_{18}|\approx0.43(5)$.

This estimate however overestimates the fidelity, since most of the signal in the experiment is actually lost due to imperfect polarization. Indeed, an imperfectly polarized state gives rise to similar oscillations, but with a smaller amplitude. To obtain an estimate of this effect, we can compare the difference signal amplitude for a simple NV-only experiment (such as a Ramsey modulation experiment) with the signal amplitude obtained in the entangling measurement. We then find $2|\rho_{18}|=0.10(2)$ indicating that, when accounting for imperfect polarization, despite creating the desired coherence, no entanglement is generated. 

\end{document}